# Climate sensitivity of Earth to solar irradiance: update

David H. Douglass*, B. David Clader, and Robert S. Knox

Department of Physics and Astronomy, University of Rochester
Rochester, NY 14627-0171

## Abstract

This paper is a continuation of a study by Douglass and Clader. We extend the analysis through December 2003 using the latest updates of the observational temperature and solar irradiance data sets in addition to a new volcano proxy data set. We have re-determined the solar effect on the temperature from satellite measurements of the solar irradiance and the temperature of the lower troposphere the sensitivity to solar irradiance. This re-analysis calculates two newly recognized dynamic and non-radiative flux factors which must be applied to the observed sensitivity. The sensitivity is about twice that expected from a no-feedback Stefan-Boltzmann radiation balance model, which implies positive feedback. The sensitivity to volcano forcing is also determined. Preliminary results indicate that negative feedback is present in this case. Response times of fractions of a year are found for both solar and volcano forcing. We note that climate models generally assume relaxation times of 5 to 10 years and we comment on the consequences of this large disparity. We also have determined a linear trend in the data.

| Abbreviations: | | |
|---|---|---|
| | AOD | Aerosol optical depth |
| | MSU | Microwave Sounding Unit |
| | SST | Sea surface temperature |
| | TLT | Name of MSU lower tropospheric data set |

*Corresponding author. douglass@pas.rochester.edu

## 1. Introduction

The importance of solar irradiance *I* and its influence on the climate of Earth has been discussed by Lean and Rind [1], White *et al*. [2], Baliunas and Soon [3], Reid [4], Crowley [5], and others. In particular, these authors recognized that the question of the sensitivity of the global-average surface temperature response of the Earth to changes in the Sun's irradiance was one of the key questions in the study of climate variability. Two of the present authors [6] determined the solar effect on Earth's temperature from satellite measurements. This paper updates their work on the basis of new observational and theoretical information.

The effect of changes in solar *I* on Earth's surface temperature *T* is smaller and can be estimated from simple radiative equilibrium models without feedback as $\Delta T/T = \Delta I/4I$. We take the average *I* as the solar constant 1365 W/m$^2$ and the average *T* as 288 K [7]. For a change $\Delta I$ ~ 1 W/m$^2$ [comparable to estimates of the amplitude of the '11 year' sunspot period], one therefore estimates $\Delta T$ ~ 0.05 K. Our measurements yield a value of about 0.10 K. This value, however, is not negligible compared to some estimates of anthropogenic effects, which are usually measured in hundreds of mK per decade. The framework for a quantitative discussion is developed below.

Models of Earth's climate system generally assume that there is a forcing $\Delta F$ (volcano, solar, CO2, etc.) that causes a change $\Delta T$ in the mean temperature of Earth's surface. In equilibrium, the relation between these is

$$\Delta T = \lambda \Delta \Phi, \qquad (1)$$

where $\lambda$ is the climate sensitivity and where $\Delta F$ is defined as an equivalent change of non-reflected solar flux averaged over Earth and referred to the "top of the atmosphere." The goal of most investigations is to determine the values of $\Delta F$ and $\lambda$ for the particular climate process under consideration. We report the determination of the observed value for solar forcing by the following method. Using multiple regression (next Section) we determine the irradiance constant $k$ defined by

$$\Delta T = k\Delta I. \tag{2}$$

The forcing $\Delta F$ due to $\Delta I$ is given by

$$\Delta F = [(1 - \alpha)/4]\, \Delta I, \tag{3}$$

which is obtained by averaging $\Delta I$ over the whole surface of the earth and allowing for a fraction (albedo $\alpha$) to be reflected away. The climate sensitivity $\lambda$ is thus given by

$$\lambda = [4/(1 - \alpha)]\, k. \tag{4}$$

## 2. Data and analysis

We use the MSU TLT lower troposphere temperature anomaly data [8].
Since 1979, satellite measurements of $I$ showing three solar activity cycles are available (Fröhlich and Lean [9] and updates) as well as lower troposphere measurements of $T$ anomalies (Christy *et al*. [8] and updates). From these two data sets we determine $k$. We determine the effect of other various geophysical phenomena by multiple regression analysis on the MSU data where a predictor $C$ for $T$ is assumed to be of the form

$$C = k_1\boldsymbol{S} + k_2\boldsymbol{V} + k_3\boldsymbol{I} + k_4\boldsymbol{L} + b\,. \tag{5}$$

We choose four predictor variables, namely El Niño, Volcano effects, Solar irradiance, and an arbitrary linear term:

1. ***S***: El Niño effects are modeled using the sea surface data SST from region 3.4 [10]. A lag time of 6 months gives the highest correlation.

2. ***V***: Volcano effects are modeled with the atmospheric optical density (AOD) [11]. We find a lag of 3 months.

3. ***I*** : Solar irradiance data of Fröhlich and Lean [9] and Fröhlich [12] are used. A lag of 3 months is found.

4. ***L***: represents a linear term, and a constant ***b***, to be discussed below.

Figure 1(a) shows the Solar irradiance ***I***, in which one clearly sees solar activity cycles 21, 22 and 23. Because the solar effect ***I*** is weaker by a factor of 10 than that of ***S*** and ***V***, we first do a regression analysis on *T* with only ***S***, ***V***, and ***L***. The resulting residuals are shown also in Figure 1(a) where a signal similar to the solar signal can be clearly seen. The autocorrelation functions are shown in Figure 1(b). The autocorrelation of the solar irradiance shows the expected cosine behavior and at a period of 10.1 years, which is the present value for the sunspot cycle. The autocorrelation of the *T* residuals and the cross correlation both show the same period.

Figure 2 shows the results of the full regression analysis. The *T* data and the predictor ***C*** are shown in the top plot. The contribution of each predictor variable is shown below. ***S*** and ***V*** are plotted together with ***S*** translated by 6 months and ***V*** by 3 months. The ***I*** and ***L*** plots and the residuals are shown lower in the figure. It is especially noted that no averaging was done before the regression analysis. The numerical results of the regression analysis and other associated quantities are shown in Table 1, where the first row gives the values of the coefficients and their standard error. The fraction of the total variance accounted for by the predictor variables is given by the coefficient $R^2$ which we determine to be 0.91.

# 3. Results and discussion

## 3.1. Collinearity

Santer *et al*. [13] have questioned the validity of regression analysis on the satellite data because large El Niño events occurred at the same time as the two volcanos which resulted in a correlation of the order of 0.4 to 0.5. They claim that such 'high' correlations indicate collinearity that can adversely affect any regression analyses such as reported here. This assertion of volcano effect on the regression coefficient is refuted by truncation experiments [6] where the coefficients were essentially unchanged by removing the Mt. Pinatubo volcano in the first truncation and El Chichón in the second. In addition, Belsley [14] has devised statistical tests to determine the presence of degrading or harmful collinearity among regression variables. Douglass *et al*. [15] have used these tests on this data to show that the regression coefficients used here have neither degrading nor harmful collinearity.

## 3.2. Solar Sensitivity

The sensitivity coefficient $k$ for solar irradiance is in fact the regression coefficient $k_3$ found above:

$$k = 0.10 \pm 0.02 \ \ \mathrm{K/(W/m^2)}. \tag{6}$$

This is the only determination of this sensitivity parameter based upon a globally complete tropospheric temperature data set. This measurement is for decadal time scales. We now calculate the value of $\lambda$ using Eq. (3) and the generally accepted value of the albedo $\alpha$ = 0.30 [7]:

$$\lambda = 0.63 \pm 0.13 \ \ \mathrm{K/(W/m^2)}. \tag{7}$$

In standard climatology theory [16], $\lambda$ depends on an intrinsic $\lambda_0$ and a gain $g$. The gain $g$ arises from processes with feedback $f$:

$$\lambda = g\lambda_0; \quad g = \frac{1}{1-f}. \tag{8a,b}$$

Rind and Lacis [17] estimate the intrinsic sensitivity as $\lambda_0$ = 0.30 K/(W/m$^2$), which has been adopted by the Intergovernmental Panel on Climate Change [18]. We therefore calculate

$$g = \frac{0.63 \pm 0.13}{0.30} = 2.1 \pm 0.4; \quad f = 0.52 \begin{array}{l} +0.08 \\ -0.12 \end{array}. \tag{9a,b}$$

This value of $f$ is consistent with that from positive water vapor feedback [19] and the delayed oscillator process proposed by White *et al*. [20].

### 3.3 Trend line in the *T* data

Whether or not $T$ shows a trend is one of the questions currently of interest. This study accounts for three of the natural effects [$\boldsymbol{S}$, $\boldsymbol{V}$, and $\boldsymbol{I}$] that obscure the observation of any underlying trend line. The data trend line for the satellite $T$ data is computed and published every month [8] and is often quoted as representing the linear trend of the data. The value of this statistic is not a reliable constant [8]. This is because the effect of $\boldsymbol{V}$ is negative and that of $\boldsymbol{S}$ can be positive (El Niño) or negative (La Niña). We show that the solar influence also affects this statistic and is negative. During this time period (1979-2003) there is an underlying decrease in the irradiance of the sun which causes a decrease in the trend of –10.6 mK/decade. We believe that the sought-for trend in the data is the coefficient of the linear term from our regression analysis. Its value is +76 ± 10 mK/decade.

**3.4. Response time and adjustments of $\lambda_0$.**

For simplicity, two factors have been omitted from the foregoing discussion of sensitivities. A more precise expression for the sensitivity to solar forcing is

$$\lambda = \lambda_0 \cdot \frac{1}{\sqrt{1+(\omega\tau)^2}} \cdot \frac{1}{1-\gamma} \cdot g\,, \tag{10}$$

where the first factor after $\lambda_0$ is the dynamical factor resulting from the frequency dependence of an assumed sinusoidal forcing with angular frequency $\omega$ and a system relaxation time $\tau$ (see, *e. g.*, [21]). The next factor is a correction to the standard Stefan-Boltzmann sensitivity resulting from non-radiative processes [22]. The last factor is the gain, as discussed earlier.

Our measured value for the response time $\tau$ of a few months is at variance with *tens of years* estimated in some energy-balance models involving the mixed-layer of the ocean. For example, Wigley and Raper [23] predict that the sunspot cycle signal would be attenuated to values of 0.02-0.03K, which is about 30% of what we observe. We suggest that this difference is due to their assumption of longer response times. Douglass *et al.* [21] in a study of the climate response to the annual solar forcing also found response times of the order of months. With a relaxation time of 3 months and a forcing period of 10.1 years, the dynamical factor becomes

$$p = \frac{1}{\sqrt{1+(\omega\tau)^2}} = \frac{1}{\sqrt{1+(2\pi\cdot 3/121)^2}} = \frac{1}{\sqrt{1+(0.15)^2}} = 0.988 \tag{11}$$

causing very little reduction. If, on the other hand, the relaxation time were 5 years, $p$ would be 0.31, a considerable reduction factor. The assumption of a long relaxation time, which is appropriate for deep-ocean processes, appears to have been made in most previous estimates of the sensitivity amplitude for 1-to-11 year periodic solar forcing. Our empirical value suggests that this assumption is not appropriate in this case. We note that Wu and North [24] in a study of seven different General Climate Models stated "…[r]elaxation times of about two to eight years".

The correction factor $1/(1-\gamma)$ was discovered recently during analysis of the energetics of a two-level model climate system [22]. The quantity $\gamma$ is directly proportional to the non-radiative (nr) flux from surface to atmosphere and it appears in the intrinsic sensitivity for the reason that the system is not entirely governed by Stefan-Boltzmann fluxes. It is not a feedback. Its value is of the order of 0.16 and is not strongly sensitive to model assumptions. The effect is to increase the Stefan-Boltzmann value of $\lambda$ by 1.19. The final result for $\lambda$ is

$$\begin{aligned} &\lambda(\text{theory}) = (0.30)(0.989)(1.19)g = 0.35g, \\ &\lambda(\text{observed}) = 0.63; \quad g = 0.63/0.35 = 1.8. \end{aligned} \tag{12a,b}$$

The dynamic factor and the nr-flux factors are seen not to be large corrections in this case. However, these two effects must be considered in all climate models and may be quite important in other studies.

### 3.5. Volcanos

We determine the volcano regression coefficient $k_2$ to be –2.9 K/$\mu$m with a delay of 3 months. Hansen *et al.* [25] have derived the following relationship to obtain the forcing from the atmospheric optical density (AOD):

$$\Delta F = -A \cdot \text{AOD}, \tag{13}$$

where $A$ = 21 W/m$^2$/$\mu$m. It follows that

$$\lambda_V = k_2/(-A) = -2.9/(-21) = 0.14 \text{ K/(W/m}^2\text{)}. \tag{14}$$

In order to estimate the gain from the volcano forcing we need to know the dynamic correction factor. For a periodic signal this factor is given by $p$ (Eq. 11). The volcano forcing is not a periodic function but one could approximate it by ¼ of a sine function of period 12 months. This allows an estimate of $p$ ~ 0.6. We estimate the sensitivity to volcano forcing to be

$$\begin{aligned} &\lambda_V(\text{theory}) = (0.30)(0.6)(1.19)g = 0.21g, \\ &\lambda_V(\text{observed}) = 0.14; \quad g = 0.14/0.21 = 0.67. \end{aligned} \tag{12a,b}$$

This result implies negative feedback $f$ = –0.5 for volcano forcing. This conclusion should be considered tentative until a better estimate is available for the dynamical factor $p$.

Many attempts have been made to explain the effects of volcano forcing. The difficulties that were encountered were perhaps due to the assumption of long relaxation times.

## 4. Summary

We find the climate sensitivity to the 11-year variation in solar irradiance to be about twice that expected from a no-feedback Stefan-Boltzmann radiation balance model. This gain of a factor of two implies positive feedback. The analysis of the sensitivity includes a consistent determination of the dynamic factor and a newly recognized non-radiative flux factor. The volcano forcing sensitivity is also determined and negative feedback is indicated. Response times of the order of 3 months are found for both solar and volcano forcing. A linear trend in the data having a slope of 76±10 mK/decade is found.

## Acknowledgements

This work was supported in part by the Rochester Area Community Foundation. We have had many useful discussions with Judith Lean, John Christy, Paul Knappenberger, and Patrick Michaels.

**Table 1.** Regression Coefficients, Trendlines, and Variances

| | TLT | TLT predictor | SST | AOD (volcano) | Solar irradiance | Linear term | Constant |
|---|---|---|---|---|---|---|---|
| Coefficient | | | $k_1$ | $k_2$ | $k_3$ | $10^4 k_4$ | $b$ |
| Units | | | K/K | K/$\mu$m | K/(W/m$^2$) | mK/decade | K |
| Value ± Std. error | | | 0.129 ± 0.008 | –2.9 ± 0.2 | 0.103 ± 0.017 | 76.8 ± 10.0 | –140 ± 23 |
| Variance × 100 | 3.79 | 2.30 | 15.0 | 11.9 | 1.9 | 3.2 | |
| Trend (mK/decade) | 85.6 | 85.6 | –5.3 | 24.7 | –10.6 | 76.8 | 0 |

$R^2$ = 1–var(residuals)/var(TLT) = 0.61. With high frequency noise removed from the residuals with an 11-month average $R^2$ = 0.91.

## Figure captions

1. (a) *T* residual values after ***S***, ***V***, and ***L*** are removed and the Solar Irradiance values showing very similar features. (b) Autocorrelation of *T* and ***I*** both showing a period of about 10.1 years.

2. (a) Satellite temperature anomalies. *T* [t2lt]. (b) t2lt predictor ***C*** based upon the predictor variables which are: (c) Sea surface temperature ***S*** shifted by 6 months, (d) Atmosphere Optical Depth ***V*** shifted by 3 months. (e) Solar Irradiance ***I***, and (f) Unknown linear effect ***L***. After subtraction one is left with the (g) Residuals.

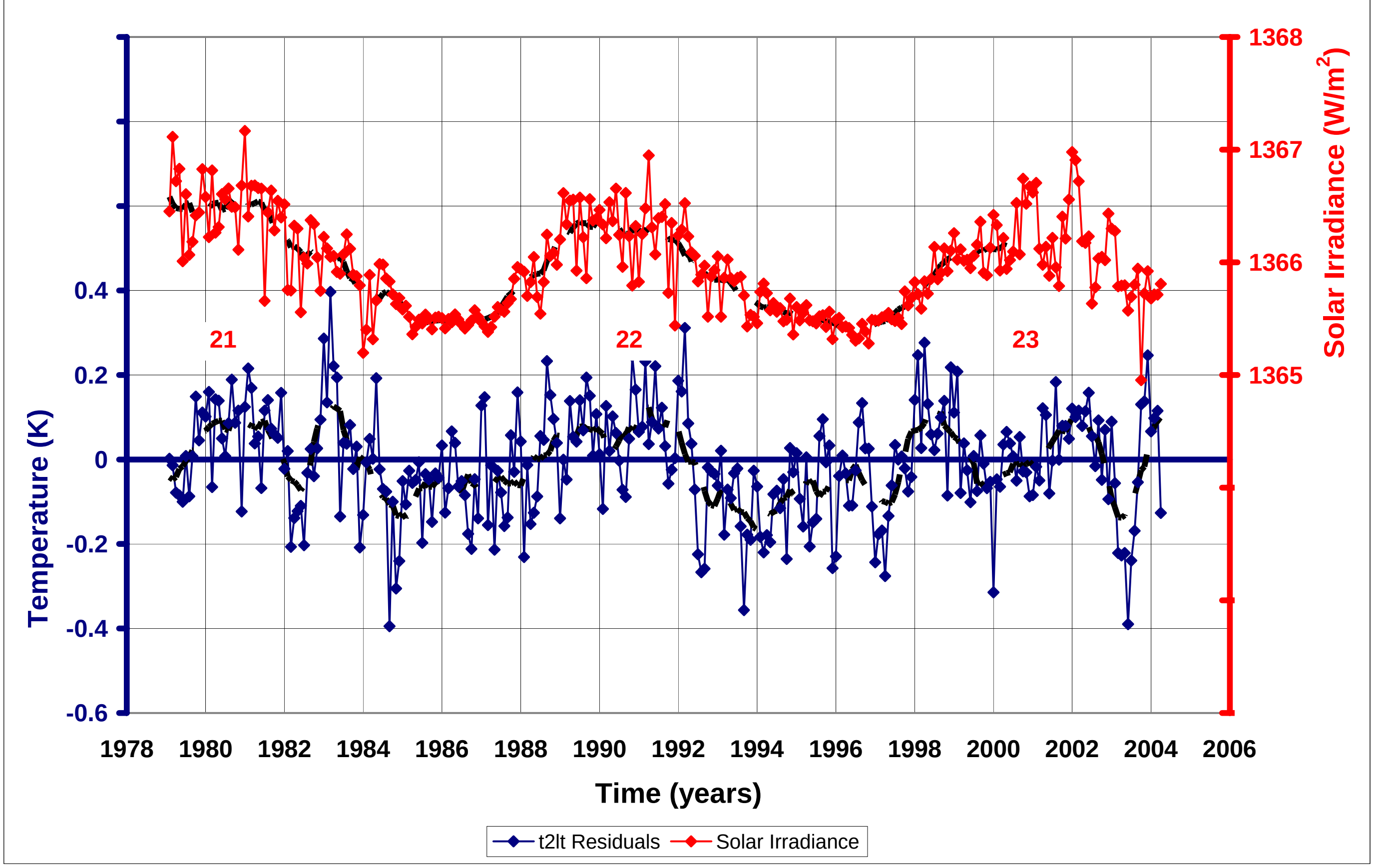

Fig 1a: Solar Irradiance and t2lt Residuals (Regression with S, V, and L Only)

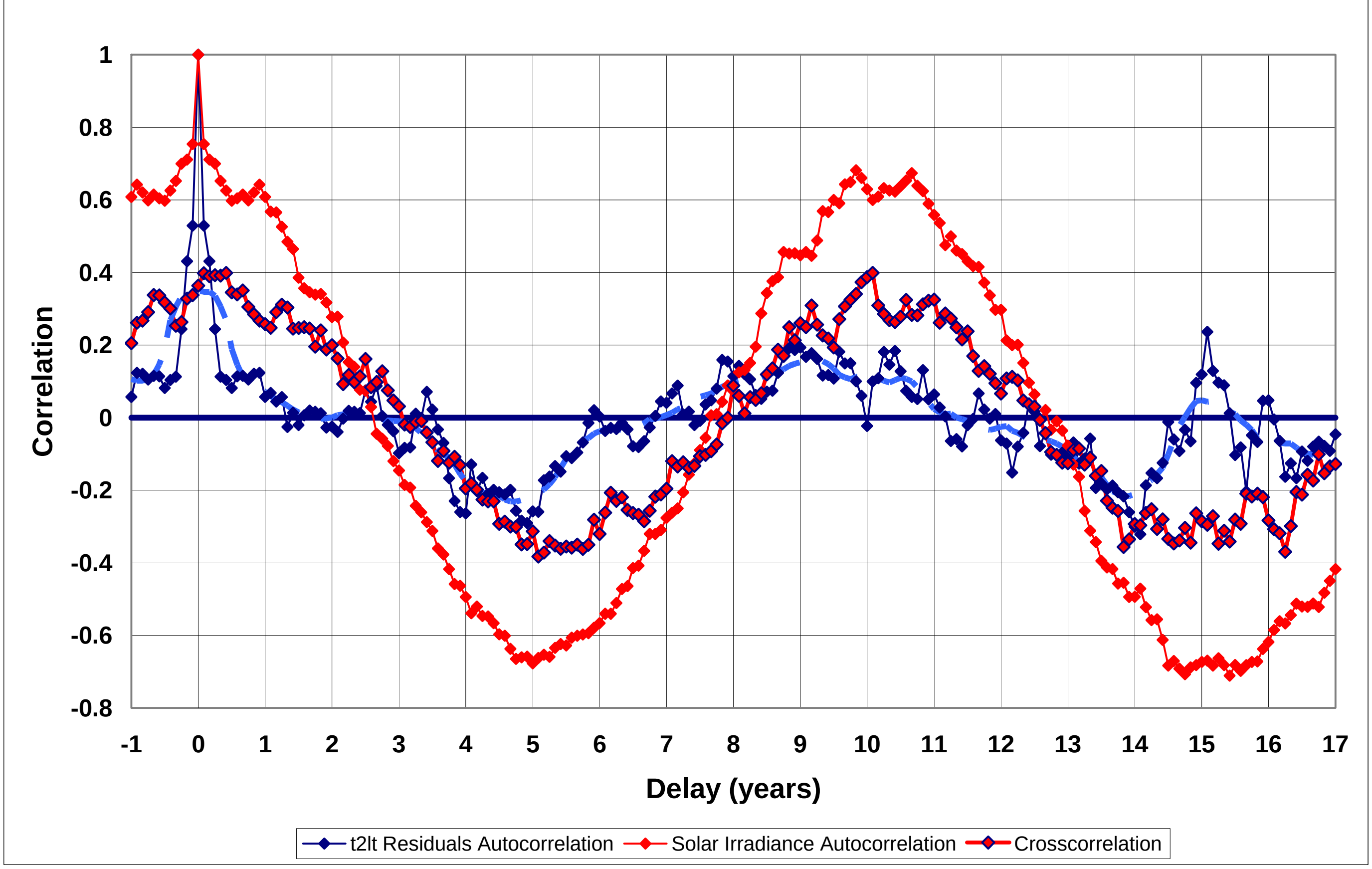

Fig 1b: Correlation Functions for Solar Irradiance and t2lt Residuals

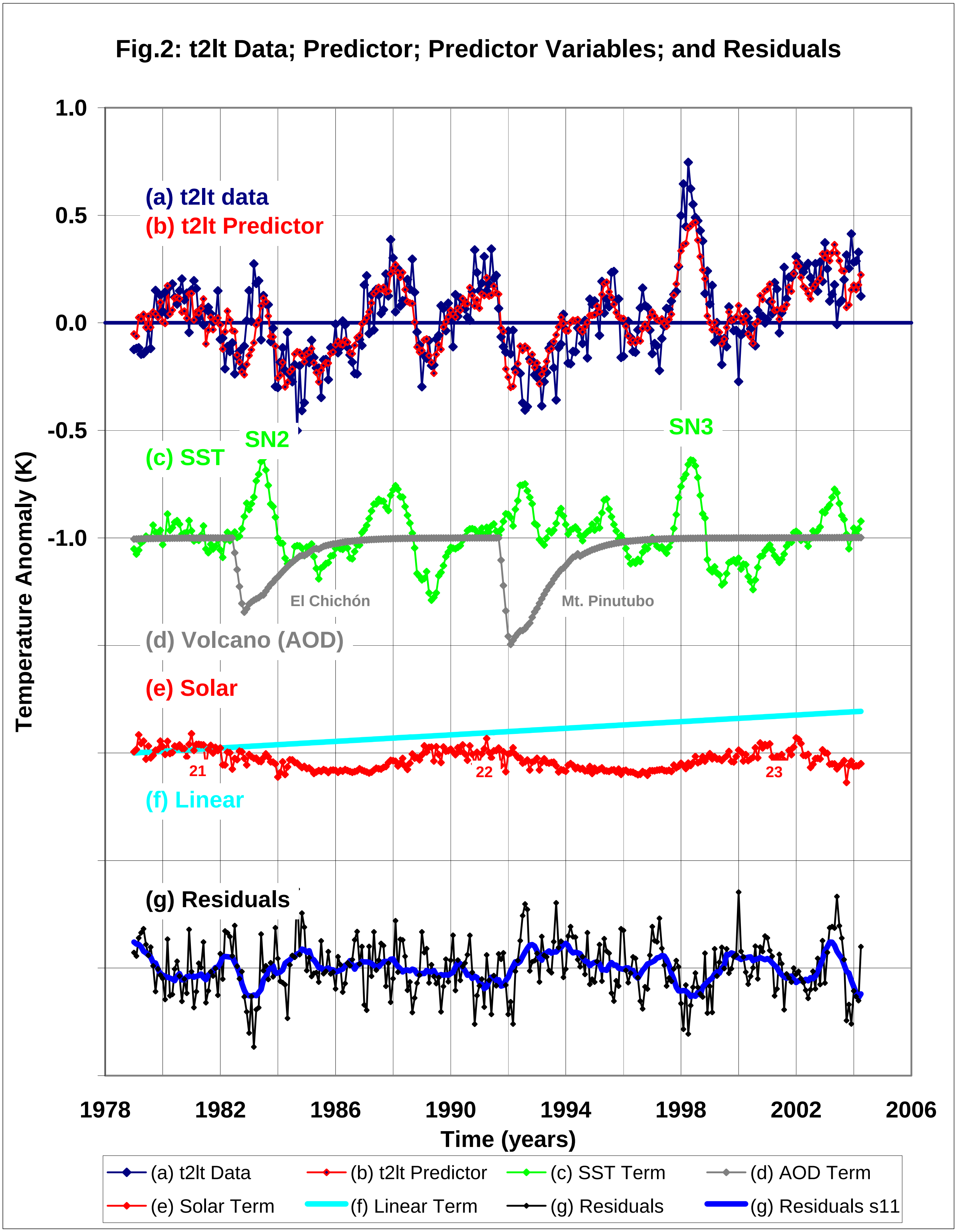

Fig.2: t2lt Data; Predictor; Predictor Variables; and Residuals
1.0
0.5
0.0
-0.5
-1.0
Temperature Anomaly (K)
(a) t2lt data
(b) t2lt Predictor
SN2
SN3
(c) SST
El Chichón
Mt. Pinutubo
(d) Volcano (AOD)
(e) Solar
21
22
23
(f) Linear
(g) Residuals
1978
1982
1986
1990
1994
1998
2002
2006
Time (years)
(a) t2lt Data
(b) t2lt Predictor
(c) SST Term
(d) AOD Term
(e) Solar Term
(f) Linear Term
(g) Residuals
(g) Residuals s11